\documentclass[manuscript=article, layout=onecolumn]{achemso}
\setkeys{acs}{articletitle = true}

\usepackage[T1]{fontenc}
\usepackage[utf8]{inputenc}
\usepackage{graphicx}
\usepackage{amsmath}
\usepackage{color}
\usepackage{comment}
\usepackage{hyperref}
\usepackage{multirow}
\usepackage{subfig}
\usepackage{bm}
\usepackage{braket}
\usepackage{siunitx}
\usepackage[normalem]{ulem}   
\usepackage[version=4]{mhchem}
\usepackage{arydshln}

\definecolor{purple}{rgb}{0.5,0,0.5}

\definecolor{corange}{rgb}{0.93, 0.53, 0.18}

\DeclareSIUnit{\kcal}{\kilo\cal}

\SectionNumbersOn 

\title{Projection-based DMRG-in-DFT embedding corrected by non-additive exchange-correlation}

\author{Enzo Monino}
\affiliation{J. Heyrovsk\'{y} Institute of Physical Chemistry, Academy of Sciences of the Czech \mbox{Republic, v.v.i.}, Dolej\v{s}kova 3, 18223 Prague 8, Czech Republic}
\altaffiliation{Contributed equally.}

\author{Daria Drwal}
\affiliation{Institute of Physics, Lodz University of Technology, \mbox{ul.\ Wolczanska 217/221, 93-005 Lodz, Poland}}
\altaffiliation{Contributed equally.}

\author{Pavel Beran}
\affiliation{J. Heyrovsk\'{y} Institute of Physical Chemistry, Academy of Sciences of the Czech \mbox{Republic, v.v.i.}, Dolej\v{s}kova 3, 18223 Prague 8, Czech Republic}
\alsoaffiliation{Faculty of Mathematics and Physics, Charles University, Prague, Czech Republic}

\author{Micha{\l} Hapka}
\affiliation{University of Warsaw, Faculty of Chemistry, ul.\ L.\ Pasteura 1, 02-093 Warsaw, Poland}

\author{Libor Veis}
\email{libor.veis@jh-inst.cas.cz}
\affiliation{J. Heyrovsk\'{y} Institute of Physical Chemistry, Academy of Sciences of the Czech \mbox{Republic, v.v.i.}, Dolej\v{s}kova 3, 18223 Prague 8, Czech Republic}

\author{Katarzyna Pernal}
\affiliation{Institute of Physics, Lodz University of Technology, \mbox{ul.\ Wolczanska 217/221, 93-005 Lodz, Poland}}
\email{pernalk@gmail.com}


\begin{document}

\begin{abstract}
The projection-based wave function (WF)-in-DFT embedding enables an
efficient description of both the energetics and properties of large
and complex chemical systems, with accuracy exceeding that of pure
DFT.
Recently, we have proposed using the density matrix renormalization group (DMRG) as the WF method for molecules containing strongly correlated fragments [Beran, P. et al. \textit{J. Phys. Chem. Lett}. 2023, 14, 3, 716–722]. In this work, we demonstrate that the accuracy of the DMRG-in-DFT approach is primarily limited by the approximate treatment of the coupling between the active component and its environment through nonadditive exchange-correlation functionals. To address this issue, we combine exact exchange to reduce the nonadditive exchange error with a multireference adiabatic connection (AC) scheme to recover nonadditive correlation. The performance of the improved DMRG-in-DFT embedding is illustrated on two prototypical strongly correlated systems: the dissociation of the H$_{20}$ chain and the cleavage of a triple CN bond in propionitrile.
\end{abstract}

\maketitle

\section{Introduction}

Strong electron correlation underpins many essential phenomena in chemistry, influencing bond dissociation, open-shell and excited-state electronic structures, and catalytic reactivity.~\cite{lyakh2012multireference, Szalay2011} Despite its fundamental importance, accurately and efficiently describing strongly correlated systems remains a major challenge in quantum chemistry. In principle, full configuration interaction (FCI) provides an exact solution to the electronic Schrödinger equation within a given one-particle basis, but its exponential scaling makes it impractical for all but the smallest systems. Consequently, a variety of approximate, polynomially scaling wave function (WF) methods have been developed, each systematically improvable toward the FCI limit. For systems dominated by weak correlation, the coupled-cluster (CC) framework~\cite{Bartlett2007} remains the method of choice. In contrast, molecules exhibiting strong (static) correlation, such as transition-metal complexes or those undergoing bond dissociation, are more appropriately treated  within the complete active space (CAS) framework~\cite{roos1987complete}, which forms the conceptual basis for this work.

The complete active space self-consistent field (CASSCF) method~\cite{Roos1980} combines a FCI treatment within a chosen active orbital space with orbital optimization, providing the standard starting point for multireference (MR) calculations. To recover the missing dynamical correlation, post-SCF approaches such as complete active space second-order perturbation theory (CASPT2)~\cite{CASPT2}, $n$-electron valence state perturbation theory (NEVPT2)~\cite{NEVPT2}, multireference adiabatic connection (AC)~\cite{Pernal2018,pastorczak2018correlation} or multireference configuration interaction (MRCI)~\cite{Szalay2011} are employed. However, all of these methods are fundamentally limited by the exponential scaling of the FCI problem within the active space, restricting practical CAS sizes to maximally 20 orbitals. 

To address this limitation, several approximate FCI solvers have been introduced. Among them, the density matrix renormalization group (DMRG) method~\cite{White1992} has become one of the most powerful techniques for treating strong correlation. Since its introduction into quantum chemistry~\cite{White1999}, DMRG has proven capable of accurately describing systems with dozens of active orbitals.~\cite{chan_review,Szalay2015,yanai_review,reiher_perspective} This success has led to the development of a number of post-DMRG approaches designed to recover out-of-CAF dynamical correlation.~\cite{Cheng2022}.
Despite the availability of methods that allow for an efficient treatment of dynamic correlation in large active spaces,\cite{DMRG-AC0} wavefunction-based approaches remain computationally demanding for large systems. In contrast, density functional theory (DFT) provides a computationally efficient framework applicable to very large molecules but suffers from its approximate exchange–correlation functionals and single-reference character, which limit its reliability for strongly correlated problems.~\cite{burke2012}

A promising route toward combining the accuracy of WF-based methods with the efficiency of DFT lies in quantum embedding frameworks.~\cite{Jones2020} These approaches exploit the locality of electronic interactions by partitioning a molecular system into an active region treated with a high-level method and an environment treated with a lower-level one.~\cite{sun2016quantum,Jones2020} Neugebauer, Reiher, and co-workers introduced the first DMRG-in-DFT embedding scheme based on the frozen density embedding (FDE) formalism.~\cite{Dresselhaus2015} However, due to the approximate nature of the nonadditive kinetic potential (NAKP), their proof-of-concept studies were limited to systems in which the active region was not covalently bonded to its environment.

The projection-based DFT (PB-DFT) embedding method~\cite{Manby2012,goodpaster2014accurate} overcomes the limitations of FDE framework by enforcing orthogonality between the occupied orbitals of the subsystems through a level-shift projection operator, thereby removing the need for an approximate NAKP. Related parameter-free formulations have also been proposed.~\cite{Khait2012,Graham2020} The PB-DFT approach is formally exact in the sense that when both subsystems are described at the same DFT level, their combined energy reproduces the total energy of the full system. Its true power, however, emerges in WF-in-DFT embedding schemes, where a high-level wave function method is applied to the active region while the surrounding environment is treated efficiently with DFT.

Building on the demonstrated success of PB-DFT embedding across diverse applications, including transition-metal catalysis, enzymatic reactivity, and electrolyte decomposition~\cite{Lee2019,puccembedding,Waldrop2021,graham2022huzinaga}, and leveraging the proven capability of DMRG to capture strong static correlation, we have previously developed the projection-based DMRG-in-DFT embedding framework.~\cite{Beran2023} This method can offer an accurate and computationally efficient approach for describing extended molecular systems with strongly correlated fragments within a multiscale electronic structure framework. Alternative multireference PB-DFT embedding approaches comprise CASSCF-in-DFT \cite{deLimaBatista2017}, or CASPT2-in-DFT \cite{Graham2020}.

We note that complementary embedding strategies include, for example, self-energy embedding theory (SEET)~\cite{Lan2015} and the subsystem embedding sub-algebra formalism~\cite{Kowalski2018}, which underpin recent developments in active-space coupled-cluster downfolding techniques.~\cite{Bauman2022}

Goodpaster et al.~\cite{goodpaster2014accurate} analyzed the sources of error in projection-based (PB) WF-in-DFT embedding and identified the dominant contribution as the nonadditive exchange–correlation (XC) error arising from the approximate nature of DFT functionals. To mitigate this problem, they proposed replacing the DFT-based nonadditive XC energy with one evaluated at the second-order Møller–Plesset (MP2) level. In related work, Bensberg and Neugebauer~\cite{Bensberg2020} demonstrated, using the Cr(CO)$_6$ complex as a representative transition-metal system, that the apparent accuracy of DFT-in-DFT embedding often stems from a delicate balance of error cancellation. They emphasized that simply substituting the DFT description of the active subsystem with a high-level WF treatment does not necessarily improve reaction energies or barriers unless the interaction energy between subsystems is also refined. This issue is expected to become even more pronounced for systems with strongly correlated active regions, where nonadditive XC errors can be substantial. In this work, we address this challenge and improve the DMRG-in-DFT method by introducing corrections to the nonadditive exchange and correlation energies, with the latter derived from the genuinely multireference adiabatic connection (AC) framework.\cite{Pernal2018, pastorczak2018correlation}

The paper is organized as follows. In Section~\ref{section_theory}, we briefly summarize the DMRG-in-DFT embedding framework and outline the computation of non-additive exchange–correlation corrections using the multireference AC approach.~\cite{Pernal2018, pastorczak2018correlation} Section~\ref{section_comp_details} presents the computational details of the benchmark calculations, and the corresponding results are discussed in Section~\ref{section_results}. Finally, conclusions are given in Section~\ref{section_conclusions}.

\section{Theory}
\label{section_theory}
In the projection-based DMRG-in-DFT \cite{Beran2023} the total energy of a composite system AB reading
\begin{align}
E_{\text{DMRG-in-DFT}}  &  =E_{\text{DMRG}}[\Psi^{ \rm A}] 
+ E_{\text{DFT}}[\gamma^{\rm A}+\gamma^{\rm B}] - E_{\text{DFT}}[\gamma^{\rm A}]\nonumber \\
& + \text{tr}\left[ (\gamma_{\text{emb}}^{\rm A}-\gamma^{\rm A})
\upsilon_{\text{emb}}[\gamma^{\rm A},\gamma^{\rm B}]\right]  \label{total_E}
\end{align}
is given in terms of density matrices of the subsystems, $\gamma^{\rm A}$ and
$\gamma^{\rm B}$, following from KS-DFT\ computation carried out in the first step
of the embedding procedure. These density matrices are defined by $N$
orthogonal spinorbitals $\left\{ \varphi_{i}\right\}_{i=1...N}$, 
$N$ indicating
a number of electrons in the composite system, assigned to one of the subsystems
comprising $N_{X}$ ($X=\rm A,\rm B$) electrons
\begin{align}
\gamma^{\rm A}(x,x^{\prime})  &  =\sum_{i\in \rm A}^{N_{\rm A}}\varphi_{i}(x)\varphi
_{i}(x^{\prime})\label{gamA}\\
\gamma^{\rm B}(x,x^{\prime})  &  =\sum_{i\in \rm B}^{N_{\rm B}}\varphi_{i}(x)\varphi
_{i}(x^{\prime}) \label{gamB}%
\end{align}
The third ingredient of the DMRG-in-DFT energy is the
wavefunction $\Psi^{A}$ and the corresponding energy $E_{\text{DMRG}}%
[\Psi^{\rm A}]$ computed as 
\begin{equation}
E_{\text{DMRG}}[\Psi^{\rm A}]=\left\langle \left.  \Psi^{\rm A}\right\vert \sum_{\sigma}\sum_{pq}h_{pq}\hat{a}_{p_{\sigma}%
}^{\dagger}\hat{a}_{q_{\sigma}}+\frac{1}{2}\sum_{\sigma\sigma^{\prime}}%
\sum_{pqrs}\left\langle pq|rs\right\rangle \hat{a}_{p_{\sigma}}^{\dagger
}\hat{a}_{q_{\sigma^{\prime}}}^{\dagger}\hat{a}_{s_{\sigma^{\prime}}}\hat
{a}_{r_{\sigma}}\left\vert \Psi^{\rm A}\right.  \right\rangle
\end{equation}
where $p,q,r,s$ are indices of general orbitals, $p_\sigma$ denotes a spinorbital.
The one-electron Hamiltonian operator,
\begin{equation}
\hat{h} = -\frac{1}{2} \Delta - \sum_I \frac{Z_I}{|\boldsymbol{r}-\boldsymbol{R}_I|},
\label{h}
\end{equation}
contains the kinetic energy and electron–nucleus attraction terms, where the summation extends over all nuclei \( I \) in the composite system.

The wavefunction $\Psi^{\rm A}$ results from a DMRG\ calculation on the subsystem A. 
It is carried out with the
Hamiltonian $\hat{H}^{\rm A}$, which includes electron-electron interactions within the
subsystem A and interaction energy of electrons with all nuclei in the composite system, and an embedding potential $\hat{\upsilon}^{\text{emb}}$, accounting for other than electron-nuclei interactions of the subsystems.
In the language of second quantization, \(\hat{H}^{\rm A}\) 
acts in the  space 
spanned by states constructed from spinorbitals belonging to the subset assigned to A
\begin{equation}
\hat{H}^{\rm A}=\sum_{\sigma}\sum_{pq\in \rm A}\left(  h_{pq}+\upsilon_{pq}%
^{\text{emb}}\right)  \hat{a}_{p_{\sigma}}^{\dagger}%
\hat{a}_{q_{\sigma}}+\frac{1}{2}\sum_{\sigma\sigma^{\prime}}\sum_{pqrs\in
\rm A}\left\langle pq|rs\right\rangle \hat{a}_{p_{\sigma}}^{\dagger}\hat
{a}_{q_{\sigma^{\prime}}}^{\dagger}\hat{a}_{s_{\sigma^{\prime}}}\hat
{a}_{r_{\sigma}} \label{HA}%
\end{equation}
The local embedding potential $\hat{\upsilon
}^{\text{emb}}$ is a bifunctional depending on density matrices of both
subsystems and it is defined as a sum of Hartree (H), exchange (x), and
correlation (c) interaction potentials. It can be written as
\begin{equation}
\hat{\upsilon}^{\rm emb}[\gamma^{\rm A},\gamma^{\rm B}](\mathbf{r})=\int\frac{\rho
^{\rm B}(\mathbf{r}^{\prime})}{|\mathbf{r-r}^{\prime}|}\ \text{d}\mathbf{r}%
^{\prime}+\hat{\upsilon}_{xc}[\gamma^{\rm A}+\gamma^{\rm B}](\mathbf{r})-\hat
{\upsilon}_{xc}[\gamma^{\rm A}](\mathbf{r})
\end{equation}
where the first term represents the Hartree potential generated by the electron 
density of subsystem B, $\rho^{\rm B}(\mathbf{r})$. Notice that in a representation of general orbitals $\{ \varphi_p(\mathbf{r}) \}$, electron density is determined by a density matrix $\gamma$
\begin{equation}
\rho(\mathbf{r}) = \sum_{pq} \sum_\sigma \gamma_{p_\sigma q_\sigma} \ \varphi_p(\mathbf{r}) \varphi_q(\mathbf{r}) 
\end{equation}
The operator $\hat{v}_{xc}$ denotes the sum of exchange and correlation potentials, 
$\hat{v}_{xc} = \hat{v}_{x} + \hat{v}_{c}$. These potentials are typically derived from 
the exchange-correlation functional \(E_{xc}[\rho]\) employed in the initial KS-DFT 
calculation of the embedding procedure, according to
\begin{equation}
\hat{v}_{xc}[\rho] = \frac{\delta E_{xc}[\rho]}{\delta \rho}
\end{equation}

The last term of the $E_{\text{DMRG-in-DFT}}$ expression, Eq.~(\ref{total_E}),
involving the density-matrix difference $\gamma_{\text{emb}}^{\rm A}-\gamma^{\rm A}$,
accounts for the lack of self-consistency between the density matrix used to
construct the embedding potential and the one resulting from the
DMRG\ computation. A fully self-consistent procedure, denoted here as
SCF-DMRG-in-DFT, in which $\gamma^{\rm A}$ is iteratively updated in $\hat{\upsilon}^{\rm emb}$, would allow this correction to be omitted. For the
sake of compactness, this term will be dropped in all subsequent equations,
regardless of whether we refer to DMRG-in-DFT\ or SCF-DMRG-in-DFT.

In order to recast the total energy expression, Eq.~(\ref{total_E}), into an
equivalent form that explicitly separates the nonadditive contributions we aim
to correct, let us first recall that within the KS-DFT\ framework the energy
functional is given by
\begin{equation}
E^{\text{DFT}}[\rho]=T_{s}[\rho]+E_{\text{ext}}[\rho]+E_{\text{H}}%
[\rho]+E_{\text{xc}}[\rho]
\end{equation}
where $T_{s}[\rho]$ denotes the noninteracting kinetic-energy functional,
$E_{\text{ext}}[\rho]$ the interaction with the external potential,
$E_{\text{H}}[\rho]$ the Hartree energy, and $E_{\text{xc}}[\rho]$ the
exchange-correlation functional.
In KS-DFT, the noninteracting reference system is described by a set of
orthogonal spinorbitals $\{\varphi_{i}\}$ that reproduce the electron density
$\rho$. The corresponding noninteracting kinetic-energy functional is defined
explicitly in terms of the KS spinorbitals
as
\begin{equation}
T_{s}[\rho]=T_{s}[\gamma]=-\frac{1}{2}\sum_{i}^N\langle\varphi_{i}%
|\Delta|\varphi_{i}\rangle\label{Ts}%
\end{equation}
where the one-particle density matrix is given by
$\gamma(x,x^{\prime})=\sum_{i}^{N}\varphi_{i}(x)\varphi_{i}(x^{\prime})$. 
Using the orthogonality of
the orbitals assigned to subsystems $\rm A$ and $\rm B$, which define the density
matrices $\gamma^{\rm A}$ and $\gamma^{\rm B}$ (recall that these orbitals originate
from a single KS-DFT calculation of the entire system and are therefore
orthogonal), together with Eq.~(\ref{Ts}) and the linearity of the functional
$E_{\text{ext}}[\rho]$, we obtain
\begin{equation}
E_{\text{ext}}[\rho^{\rm A}+\rho^{\rm B}] =  \int \left(\rho^{\rm A}(\mathbf{r}) 
+ \rho^{\rm B}(\mathbf{r})\right) \upsilon_{\text{ext}}(\mathbf{r})  \, \text{d}\mathbf{r}
= E_{\text{ext}}[\rho^{\rm A}] +  E_{\text{ext}}[\rho^{\rm B}]
\end{equation}
so that the energy $E_{\text{DFT}}[\gamma^{\rm A}+\gamma^{\rm B}]$ appearing in
Eq.~\eqref{total_E} can be decomposed as follows:
\begin{equation}
E_{\text{DFT}}[\gamma^{\rm A}+\gamma^{\rm B}] = T_{s}[\gamma^{\rm A}] + T_{s}[\gamma^{\rm B}] 
+ E_{\text{ext}}[\gamma^{\rm A}] + E_{\text{ext}}[\gamma^{\rm B}] + E_{\text{H}}[\gamma^{\rm A}+\gamma^{\rm B}]
+ E_{\text{xc}}[\gamma^{\rm A}+\gamma^{\rm B}]
\end{equation}

Consequently, the difference between the DFT\ energy of the composite system
and that of subsystem A is expressed as the sum of the environment energy B
and the Hartree and exchange-correlation nonadditive terms. Namely,
\begin{equation}
E_{\text{DFT}}[\gamma^{\rm A}+\gamma^{\rm B}]-E_{\text{DFT}}[\gamma^{\rm A}]=E_{\text{DFT}%
}[\gamma^{\rm B}]+E_{H}^{\text{nadd}}[\gamma^{\rm A},\gamma^{\rm B}]+E_{xc}^{\text{nadd}%
}[\gamma^{\rm A},\gamma^{\rm B}]\ \ \
\end{equation}
where the nonadditive Hartree energy functional reads%
\begin{equation}
E_{H}^{\text{nadd}}[\gamma^{\rm A},\gamma^{\rm B}]\equiv E_{H}[\gamma^{\rm A}+\gamma
^{\rm B}]-E_{H}[\gamma^{\rm A}]-E_{H}[\gamma^{\rm B}]=\int\int\frac{\rho^{\rm A}%
(\mathbf{r})\rho^{\rm B}(\mathbf{r}^{\prime})}{|\mathbf{r-r}^{\prime}|}%
\ \text{d}\mathbf{r}\text{d}\mathbf{r}^{\prime}\ \ \
\end{equation}
$E_{xc}^{\text{nadd}}$ is defined as the sum of the nonadditive exchange and
correlation functionals, 
\begin{equation}
E_{xc}^{\text{nadd}}[\gamma^{\rm A},\gamma^{\rm B}]=E_{x}^{\text{DFT,nadd}}[\gamma
^{\rm A},\gamma^{\rm B}]+E_{c}^{\text{DFT,nadd}}[\gamma^{\rm A},\gamma^{\rm B}]
\end{equation}
where
\begin{equation}
E_{x}^{\text{DFT,nadd}}[\gamma^{\rm A},\gamma^{\rm B}]\equiv E_{x}[\gamma^{\rm A}%
+\gamma^{\rm B}]-E_{x}[\gamma^{\rm A}]-E_{x}[\gamma^{\rm B}]
\end{equation}
and analogously for the correlation functional $E_{c}^{\text{DFT,nadd}}$. Both
the nonadditive exchange and correlation functionals are approximate and thus
inherit the errors of the underlying exchange and correlation functionals.

With the above definitions of the nonadditive functionals, the total energy,
Eq.~(\ref{total_E}), can be rewritten in the equivalent form
\begin{align}
E_{\text{DMRG-in-DFT}} & =E_{\text{DMRG}}[\Psi^{\rm A}]+E_{\text{DFT}%
}[\gamma^{\rm B}]+E_{H}^{\text{nadd}}[\gamma^{\rm A},\gamma^{\rm B}] \nonumber \\
 & + E_{c}%
^{\text{DFT,nadd}}[\gamma^{\rm A},\gamma^{\rm B}]+E_{x}^{\text{DFT,nadd}}[\gamma
^{\rm A},\gamma^{\rm B}] \label{total_E_nadd}%
\end{align}

\noindent
If $\Psi^{\rm A}$ is obtained from a self-consistent procedure, it
may be more appropriate to evaluate the nonadditive energy terms with
$\gamma_{\text{emb}}^{\rm A}$, which leads to
\begin{align}
E_{\text{SCF-DMRG-in-DFT}} & = E_{\text{DMRG}}[\Psi^{\rm A}%
]+E_{\text{DFT}}[\gamma^{\rm B}]+E_{H}^{\text{nadd}}[\gamma_{\text{emb}}%
^{\rm A},\gamma^{\rm B}]  \nonumber \\
 & + E_{c}^{\text{DFT,nadd}}[\gamma_{\text{emb}}^{\rm A},\gamma
^{\rm B}]+E_{x}^{\text{DFT,nadd}}[\gamma_{\text{emb}}^{\rm A},\gamma^{\rm B}]
\end{align}

The nonadditive exchange and correlation energy functionals describe the
exchange and correlation between subsystems A and B. When semilocal
functionals are applied to strongly correlated systems, they introduce
significant static correlation errors. 
Our goal is to correct the nonadditive exchange-correlation contributions.

Exploiting the orthogonality of the orbitals used to construct $\gamma
_{\text{emb}}^{\rm A}$ and $\gamma^{\rm B}$, we propose to replace the nonadditive
exchange functional by its exact, nonlocal counterpart. The nonadditive
exchange correction for SCF-DMRG-in-DFT energy
is therefore defined as
\begin{equation}
\Delta_{x}^{\text{nadd}} =-E_{x}^{\text{DFT,nadd}}[\gamma_{\text{emb}}^{\rm A},\gamma^{\rm B}]
+E_{xx}^{\text{nadd}}[\gamma_{\text{emb}}^{\rm A},\gamma^{\rm B}]
\label{delta_x}
\end{equation}
The nonadditive exact exchange energy is of the form
\begin{equation}
E_{xx}^{\text{nadd}}[\gamma_{\text{emb}}^{\rm A},\gamma^{\rm B}] =-\sum_{\sigma}\sum_{p q \in
\rm A}\sum_{r s \in \rm B} \left[ \gamma_{\text{emb}}^{\rm A}\right] _{p_\sigma q_\sigma} \left[
\gamma^{\rm B}\right] _{r_\sigma s_\sigma} \langle pr|sq\rangle
\label{Del_x}%
\end{equation}
where A and B denote disjoint subsets of orbitals assigned to the fragments. 

The nonadditive correlation correction, $\Delta_{c}^{\text{nadd}}$, is
formulated by exploiting the AC0 correlation energy expression\cite{pastorczak2018correlation} and retaining
only the terms that describe inter-fragment correlation,
\begin{equation}
\Delta_{c}^{\text{nadd}} =-E_{c}^{\text{DFT,nadd}}[\gamma_{\text{emb}}^{\rm A},\gamma^{\rm B}]
+E_{c}^{\text{AC0,nadd}}[\gamma_{\text{emb}}^{\rm A},\gamma^{\rm B}]
\end{equation}
Note that the computation of $E_{c}^{\text{AC0,nadd}}$ requires both the 1-
and 2-RDMs of fragment A, although for compactness we indicate the dependence on
the 1-RDM only. A physical nonadditive correlation energy must vanish in the
limit of dissociation of the composite system into noninteracting fragments.
Accordingly, the construction of the nonadditive AC0 correlation energy is
subject to a condition
\begin{equation}
\lim_{R_{\rm AB}\rightarrow\infty} E_{c}^{\text{AC0,nadd}}[\gamma_{\text{emb}}^{\rm A},\gamma^{\rm B}]
=0\label{cond1}%
\end{equation}

Let us analyze the AC0 correlation energy expressed in terms of products of the correlation amplitudes $T^{\text{AC0}}$ and two-electron
integrals\cite{pastorczak2018correlation,guo2024spinless}:
\begin{equation}
E_{\text{corr}}^{\text{AC0}} = \sum_{pqrs} T_{pq,rs}^{\text{AC0}} \langle
pr|qs \rangle
\end{equation}
If the orbital space is partitioned into active (a), inactive (doubly
occupied, o), and virtual (v) orbitals, corresponding to fractional, 2, or 0
occupation numbers, respectively, then only the following combinations of
pairs $(pq)$ and $(rs)$ are allowed: (ao)(ao), (va)(va), (vo)(aa), (va)(aa),
(aa)(ao), (va)(ao), (vo)(ao), (vo)(vo), (va)(vo). Here, the notation (vo)(ao)
indicates that in the first pair $(pq)$, one orbital is virtual (v) and
one is inactive (o), while in the second pair $(rs)$, one orbital is
active (a) and the other is inactive (o).

A\ completion of the DMRG-in-DFT\ calculation leads to obtaining a set of
occupied orbitals localised on the subsystem B (those are the frozen
KS-DFT\ orbitals), which will be denoted by o$_\text{B}$, and a set of active orbitals assigned to the subsystem A. If
the concentric localisation procedure \cite{Claudino2019b}, or some alternative truncation scheme, has been adopted to reduce the dimension
of the orbital space for DMRG calculation, then there will also be a set of
virtual (v) orbitals. However, by construction the virtual orbitals do not correlate
with the active orbitals, consequently the $T_{\text{(va)(..)}}^{\text{AC0}}$
AC0 amplitudes are assumed to be negligible. 
We now consider which of the remaining amplitude classes, (ao$_\text{B}$)(ao$_\text{B}$), (vo$_\text{B}$)(aa), (aa)(ao$_\text{B}$), (vo$_\text{B}$)(ao$_\text{B}$) satisfy condition \eqref{cond1} and thus must be included in the nonadditive correlation energy. 
Before analysing them, one should notice that in the dissociation limit
$R_{\rm AB}\rightarrow\infty$, some active orbitals will localise on the fragment
B and their occupation numbers will vanish (they will become uncorrelated).
Denote a set of such orbitals as (a$_{\text{B}}$). The complementary set,
denoted as (a$_{\text{A}}$) will consist of the active orbitals which will
localise on the subsystem A. Now it is clear that contributions to the AC0
correlation energy from the amplitudes $T_{\text{(ao$_\text{B}$)(ao$_\text{B}$)}}^{\text{AC0}}$ and
$T_{\text{(vo$_\text{B}$)(ao$_\text{B}$)}}^{\text{AC0}}$ do not vanish in the dissociation limit, as
terms with the active orbitals belonging to a subset (a$_{\text{B}}$), namely
$T_{\text{(a}_{\text{B}}\text{o$_\text{B}$)(a}_{\text{B}}\text{o$_\text{B}$)}}^{\text{AC0}}$ and
$T_{\text{(vo$_\text{B}$)(a}_{\text{B}}\text{o$_\text{B}$)}}^{\text{AC0}}$, will remain finite (they
can be interpreted as those which describe a correlation energy for the
subsystem B),
\begin{equation}
\exists_{\substack{pq\in(\text{a}_{\text{B}}\text{o})\\rs\in(\text{a}%
_{\text{B}}\text{o})}}\ \ \wedge\ \ \exists_{\substack{pq\in(\text{vo}%
)\\rs\in(\text{a}_{\text{B}}\text{o})}}\ \ \ \lim_{R_{\rm AB}\rightarrow\infty
}T_{pq,rs}^{\text{AC0}}\neq0
\label{T1}
\end{equation}
One is therefore left with only two classes of amplitudes, 
$T_{\text{(vo$_\text{B}$)(aa)}}^{\text{AC0}}$ and $T_{\text{(aa)(ao$_\text{B}$)}}^{\text{AC0}}$. 
Consider first the case when both active indices in a pair (aa) 
correspond to orbitals localized on subsystem B. 
In this situation, the amplitudes vanish in the dissociation limit, 
as the occupancies of the orbitals $a_{\text{B}}$ approach zero, namely
\begin{equation}
\forall_{p\in(\text{a}_{\text{B}})} \ \ \lim_{R_{\rm AB}\rightarrow\infty} n_{p}=0 
\ \ \Longrightarrow \ \ 
\forall_{\substack{pq\in(\text{vo}) \\ rs\in(\text{a}_{\text{B}}\text{a}_{\text{B}})}} 
\ \wedge \ 
\forall_{\substack{pq\in(\text{a}_{\text{B}}\text{o}) \\ rs\in(\text{a}_{\text{B}}\text{a}_{\text{B}})}} 
\ \ \lim_{R_{\rm AB}\rightarrow\infty} T_{pq,rs}^{\text{AC0}}=0 
\label{T2}
\end{equation}
In all other cases, the (vo$_\text{B}$)(aa) and (aa)(ao$_\text{B}$) amplitudes involve at least one 
active orbital from a subset (a$_\text{A}$). Since orbitals o$_\text{B}$ are localized on B, these amplitudes also vanish in the dissociation limit (recall that only amplitudes with indices corresponding to orbitals localized on the same fragment remain finite). Altogether, we obtain
\begin{equation}
\forall_{\substack{pq\in(\text{vo$_\text{B}$}) \\ rs\in(\text{aa})}} 
\ \wedge \ 
\forall_{\substack{pq\in(\text{ao$_\text{B}$}) \\ rs\in(\text{aa})}} 
\ \ \lim_{R_{\rm AB}\rightarrow\infty} T_{pq,rs}^{\text{AC0}}=0 
\label{T3}
\end{equation}
%
			




The proposed nonadditive AC0 correlation energy for DMRG-in-DFT, which
satisfies the condition in Eq.(\ref{cond1}), includes therefore only two types
of amplitudes, namely
\begin{equation}
E_{c}^{\text{AC0,nadd}} = 2\sum_{\substack{\left( pq\right) \in (\text{vo}_\text{B})\\(rs) \in \text{(aa)}}} T_{pq,rs}^{\text{AC0}} \left\langle pr|qs \right\rangle
+2 \sum_{\substack{\left(pq\right) \in \text{(aa)} \\ (rs) \in (\text{ao}_\text{B}) }} 
T_{pq,rs}^{\text{AC0}} \left\langle pr|qs\right\rangle \label{nadd_c}
\end{equation}
(a factor 2 results from exploiting the symmetry $T_{pq,rs}^{\text{AC0}}=T_{rs,pq}^{\text{AC0}}$).

The total DMRG-in-DFT energy, corrected for the nonadditive exchange and correlation
contributions, is finally given by
\begin{equation}
E = E_{\text{DMRG-in-DFT}} + \Delta_{x}^{\text{nadd}} + \Delta_{c}^{\text{nadd}}
\end{equation}

To increase the computational efficiency of treating subsystem A, one may consider replacing DMRG with CASSCF within the embedding framework. 
Employing CASSCF instead of DMRG has two main implications. 
First, unlike DMRG, the CASSCF energy of subsystem A does not include dynamic correlation, which necessitates an additional evaluation of the corresponding correlation energy. 
Second, in contrast to the DMRG case with concentric localization, all active orbitals in CASSCF remain localized on fragment A and are correlated with the virtual orbitals. 
Consequently, the nonadditive correlation energy will involve a larger set of AC0 amplitude classes than those considered for DMRG-in-DFT as shown in Eq.~(\ref{nadd_c}).

To formulate a nonadditive correlation energy expression appropriate for CAS-in-DFT, we require, as before, that it satisfies the dissociation condition given in Eq.~(\ref{cond1}). 
Since a given amplitude $T_{pq,rs}^{\text{AC0}}$ vanishes unless all orbitals $p, q, r, s$ are localized on the same fragment in the dissociation limit, and because both inactive and active orbitals used in the CAS wavefunction remain localized on subsystem A in this limit, it follows that all amplitudes for which at least one index corresponds to subsystem A and at least one to subsystem B vanish upon dissociation. 
All such amplitudes are included in the computation of the nonadditive correlation energy,
\begin{equation}
E_{c}^{\text{AC0,nadd}} = \sum_{(pqrs)\in\textbf{(A-B)}_c} 
T_{pq,rs}^{\text{AC0}} \langle pr|qs\rangle,
\end{equation}
and are listed in the first column of Table~\ref{ac0_amplitudes_partitioning} [the notation $(pqrs)\in\textbf{(A-B)}_c$ indicates that a correlation amplitude $T_{pq,rs}$ belongs to a set $\textbf{(A-B)}_c$].

\begin{center}
\begin{table}[h!]
\centering
\caption{Partitioning of AC0 amplitudes into contributions to nonadditive correlation (first column) and subsystem A correlation energy (second column) in CAS-in-DFT. 
o$_\text{A}$ and o$_\text{B}$ denote inactive orbitals assigned to subsystems A and B, respectively, while a and v pertain to active (fractionally occupied, assigned to A) and virtual orbitals.
For the DMRG-in-DFT variant only two terms contribute to the nonadditive correlation energy; these are marked with an asterisk. 
}
\label{tab:corr}
\renewcommand{\arraystretch}{1.5} 
\begin{tabular}{cc}
\hline
\textbf{(A-B)}$_c$ & \textbf{A}$_c$ \\
\hline
(aa)(ao$_\text{B}$)$^*$   & (aa)(ao$_\text{A}$)  \\
\hdashline
(aa)(vo$_\text{B}$)$^*$   & (aa)(vo$_\text{A}$) \\
\hdashline
(va)(ao$_\text{B}$)   &   (va)(ao$_\text{A}$)\\
\hdashline
(va)(vo$_\text{B}$)   & (va)(vo$_\text{A}$)\\
\hdashline
(ao$_\text{B}$)(ao$_\text{B}$) & \multirow{2}{*}{(ao$_\text{A}$)(ao$_\text{A}$)} \\
(ao$_\text{B}$)(ao$_\text{A}$) &  \\
\hdashline
(vo$_\text{B}$)(ao$_\text{A}$) & \\
(vo$_\text{A}$)(ao$_\text{B}$) &  (vo$_\text{A}$)(ao$_\text{A}$)\\
(vo$_\text{B}$)(ao$_\text{B}$) &  \\
\hdashline
(vo$_\text{A}$)(vo$_\text{B}$) & (vo$_\text{A}$)(vo$_\text{A}$) \\
\hdashline
  &  (va)(aa)\\
  \hdashline
   & (va)(va)\\
\hline
\end{tabular}
\label{ac0_amplitudes_partitioning}
\end{table}
\end{center}

Amplitudes $T_{pq,rs}^{\text{AC0}}$ contributing to the dynamic correlation
energy of subsystem A are those for which none of the indices $p,q,r,s$
correspond to orbitals assigned to subsystem B.
The total CAS-in-DFT energy, corrected for the correlation energy of subsystem
A as well as the nonadditive exchange and correlation contributions, is therefore given by
\begin{equation}
E=E_{\text{CAS-in-DFT}}+\sum_{(pqrs)\in\textbf{A}_c}T_{pq,rs}^{\text{AC0}}\langle
pr|qs\rangle+\Delta_{x}^{\text{nadd}}+\Delta_{c}^{\text{nadd}}
\label{CAS}
\end{equation}
where the summation runs over all AC0 amplitudes associated with the correlation on
subsystem A, i.e.\ those which belong to a set $\textbf{A}_c$ presented in the second column of Table~\ref{ac0_amplitudes_partitioning}. 

The proposed corrections to the nonadditive correlation energy are based on the AC0 method, as it provides an efficient ab initio route for computing multireference correlation energies in large active spaces, making it particularly suitable for DMRG embedding calculations. 
Nevertheless, an analogous nonadditive correlation correction could equally well be formulated using a multireference perturbation theory method, such as CASPT2\cite{CASPT2} or NEVPT2\cite{NEVPT2}, by following the same principles as in AC0 [see Eqs.~(\ref{T1})--(\ref{T3})] to select the PT2 amplitudes\cite{guo2025approximation}, $T_{pqrs}^{\text{PT2}}$, contributing to the nonadditive correlation energy $E_{\text{c}}^{\text{PT2,nadd}}$.

Finally, we note that the exchange and correlation corrections proposed in this work
represent multireference extensions of the analogous corrections introduced by
Goodpaster et al.~\cite{goodpaster2014accurate} to reduce the CCSD(T)-in-DFT
embedding error. In particular, for a single-reference description of the active fragment A, the AC0 amplitudes appearing in Eqs.~(\ref{nadd_c}) and~(\ref{CAS}) reduce to the MP2 amplitudes $T_{ia,jb}^{\text{MP2}}$, where $i \in \rm A$, $j \in \rm B$, and $a,b$ denote virtual orbitals. In this limit, the nonadditive correlation correction becomes identical to that proposed in Ref.~\citenum{goodpaster2014accurate}. 
The nonadditive exchange energy correction introduced here, cf.\ Eq.~(\ref{delta_x}),
is formally analogous to that of Goodpaster et al., 
with the  difference that in the present work a correlated density matrix
of subsystem A is employed, whereas in Ref.~\citenum{goodpaster2014accurate}
the uncorrelated (Hartree--Fock) embedded density matrix was used.

\section{Computational details}
\label{section_comp_details}

\begin{figure}[!ht]
    \centering
    \subfloat[\label{h20}]{%
    \includegraphics[width=1\linewidth]{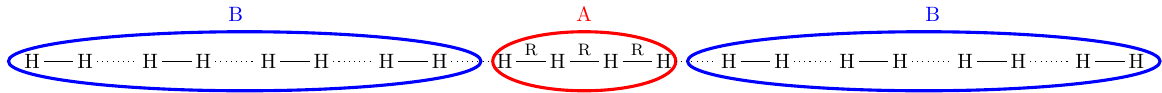}
    } \\
    \subfloat[\label{prop}]{%
    \includegraphics[width=0.8\linewidth]{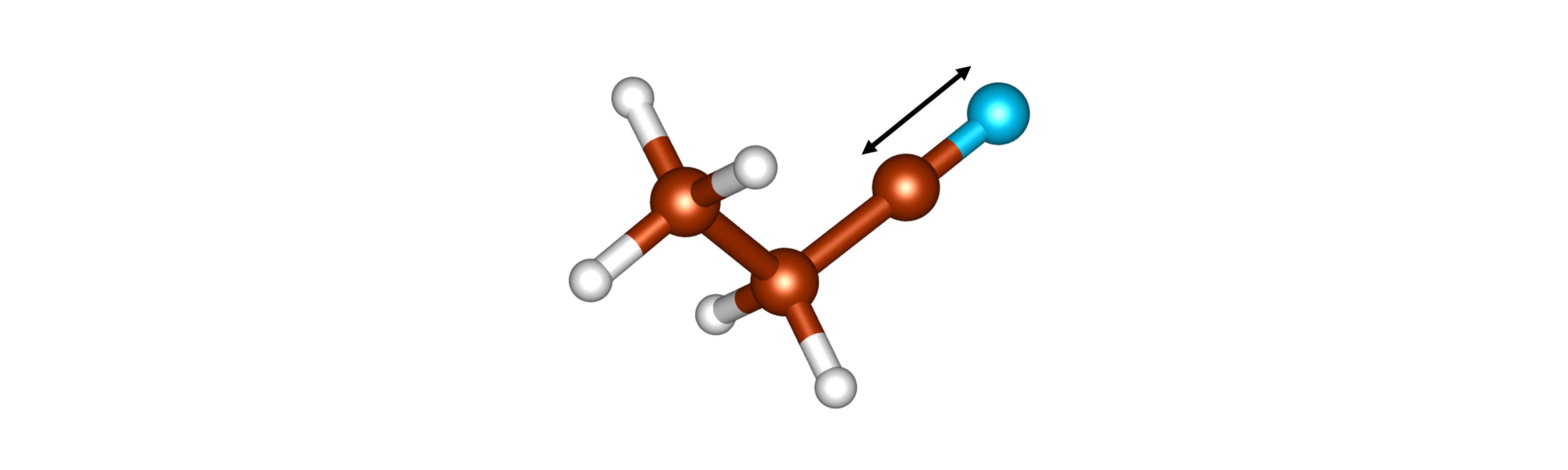}
    }
    \caption{Benchmark problems investigated in this work: (a) H$_{20}$ chain with the central active fragment (A) composed of four hydrogen atoms, where the interatomic distances ($R$) are varied, (b) triple bond stretching in propionitrile (CH$_3$CH$_2$CN) with the active fragment defined as the $-$CN group.}
    \label{systems}
\end{figure}

We applied the methodology described in the previous section to two prototypical molecular systems containing strongly correlated fragments. As a representative example, we considered a linear chain of 20 hydrogen atoms (H$_{20}$, Figure \ref{h20}), with the central active fragment consisting of four equally spaced hydrogen atoms. To study the effect of increasing multireference character, we varied the interatomic distances within this fragment from 0.7 to 2.5 \AA. The remaining hydrogen atoms form the environment, arranged as hydrogen dimers with an intra-dimer bond length of 1.0 \AA~and an inter-dimer separation of 1.4 \AA. 
The latter distance also corresponds to the separation between the hydrogen nuclei at the edges of subsystem A and the nearest hydrogen nuclei belonging to the environment. 
The distance 1.4~\AA\ is short enough so that covalent bonds between the central H$_4$ unit and the neighboring hydrogen atoms are not fully broken. As discussed below, this distance ensures strong coupling between the edge atoms of fragment~A and those of fragment~B.
To assess the effect of enlarging the active fragment, we also examined a case where the active region contains eight hydrogen atoms, the four stretched central atoms plus the two nearest hydrogen dimers (one on each side). The hydrogen chain was chosen for its one-dimensional character, as the DMRG method can provide reliable benchmark energies of near-FCI quality for both basis sets employed, namely 6-31G \cite{Hehre1972} and cc-pVDZ \cite{Dunning1989}.

The second system we investigated is propionitrile (CH$_3$CH$_2$CN, Figure~\ref{prop}), for which we adopted the molecular geometry from our previous work on DMRG-in-DFT \cite{Beran2023}. The active subsystem was defined as the $-$CN group, and by stretching the \ce{C#N} triple bond from 0.85 to 2.35 \AA, we modulated its multireference character. In the case of propionitrile, all calculations were performed using the cc-pVDZ basis set \cite{Dunning1989}. As a reference, we used DMRG energies within the frozen-core approximation, taken from Ref.~\citenum{Beran2023}.

In the following section, we compare two WF-in-DFT approaches: DMRG-in-DFT \cite{Beran2023} and CASSCF-in-DFT \cite{deLimaBatista2017}, which we denote shortly as CAS-in-DFT. The occupied orbitals, obtained from the KS-DFT computation on the entire molecule, were partitioned into active (A) and environment (B) subspaces using the SPADE procedure \cite{Claudino2019a}. For DMRG-in-DFT, a preliminary HF-in-DFT calculation was performed.
In order to maintain orthogonality between the active and environment orbitals, we employed the parameter-free approach diagonalizing the modified Fock matrix with the environment degrees of freedom projected out \cite{Khait2012}.
In the case of CASSCF-in-DFT, the original projector method \cite{Manby2012} was applied with a projector scaling parameter of $10^6$. Additionally, for DMRG-in-DFT, we employed the concentric localization (CL) technique\cite{Claudino2019b} to reduce the size of the virtual space.

All DMRG calculations, both embedded and reference, were carried out in a basis of Pipek–Mezey \cite{PipekMezey1989} split-localized molecular orbitals \cite{OlivaresAmaya2015} by means of the \textsf{MOLMPS} program \cite{Brabec2020}. Orbital ordering was optimized using the Fiedler method \cite{barcza_2011} applied to the matrix of exchange integrals \cite{OlivaresAmaya2015}, and the calculations were initialized with the CI-DEAS procedure \cite{Szalay2015, Legeza2003}. We employed the dynamical block state selection (DBSS) scheme \cite{legeza_2003a}, which adapted the bond dimension to achieve a target truncation error of $10^{-6}$. The minimum bond dimension and the fixed bond dimension used during the first warm-up sweep were both set to 500 for the embedded calculations and 1000 for the reference ones.

The CASSCF as well as DFT calculations were performed with \textsf{ORCA} package \cite{orca}. In both systems, the active space was defined as the strongly correlated orbital subspace of the active fragment, CAS(4,4) and CAS(8,8) for H$_{20}$ with four- and eight hydrogen atoms in the active fragment, respectively, and CAS(6,6) for propionitrile. The projection-based WF-in-DFT protocol was implemented in a local version of \textsf{ORCA 5.0}. For DFT calculations, we employed the B3LYP \cite{Becke1993_ExactExchange, Lee1988_LYP, Stephens1994_B3LYPuse}, and PBE \cite{PBE1996} exchange–correlation functionals.

The AC0 correlation energy corrections were computed in the \textsf{GammCor} program \cite{gammcor}. 
The amplitudes listed in Table~\ref{tab:corr} were obtained from the general
expression given, for example, in Eq.~(79) of Ref.~\citenum{guo2024spinless}.
In general, computing the AC0 amplitudes requires solving the zeroth-order
extended random phase approximation (ERPA) equations,\cite{erpa1} which involve an
uncorrelated Hamiltonian $\hat{H}^{(0)}$ constructed for a given reference
wavefunction. For the embedding schemes considered here, the one-electron
part of $\hat{H}^{(0)}$ must be modified to reflect that the active orbitals in
DMRG are optimized in the presence of the embedding potential generated by
the environment, whereas in CASSCF they are influenced by both the embedding
potential and the inactive orbitals localized on subsystem~A. The explicit
forms of the effective one-electron Hamiltonians employed for evaluating the
AC0 amplitudes, and consequently the corresponding nonadditive correlation energy
corrections in the DMRG-in-DFT and CAS-in-DFT schemes, are provided in the
Appendix.

\section{Results and discussion}
\label{section_results}

\subsection{H$_{20}$ chain}
\begin{figure}[!ht]
    \centering
    \includegraphics[width=0.6\linewidth]{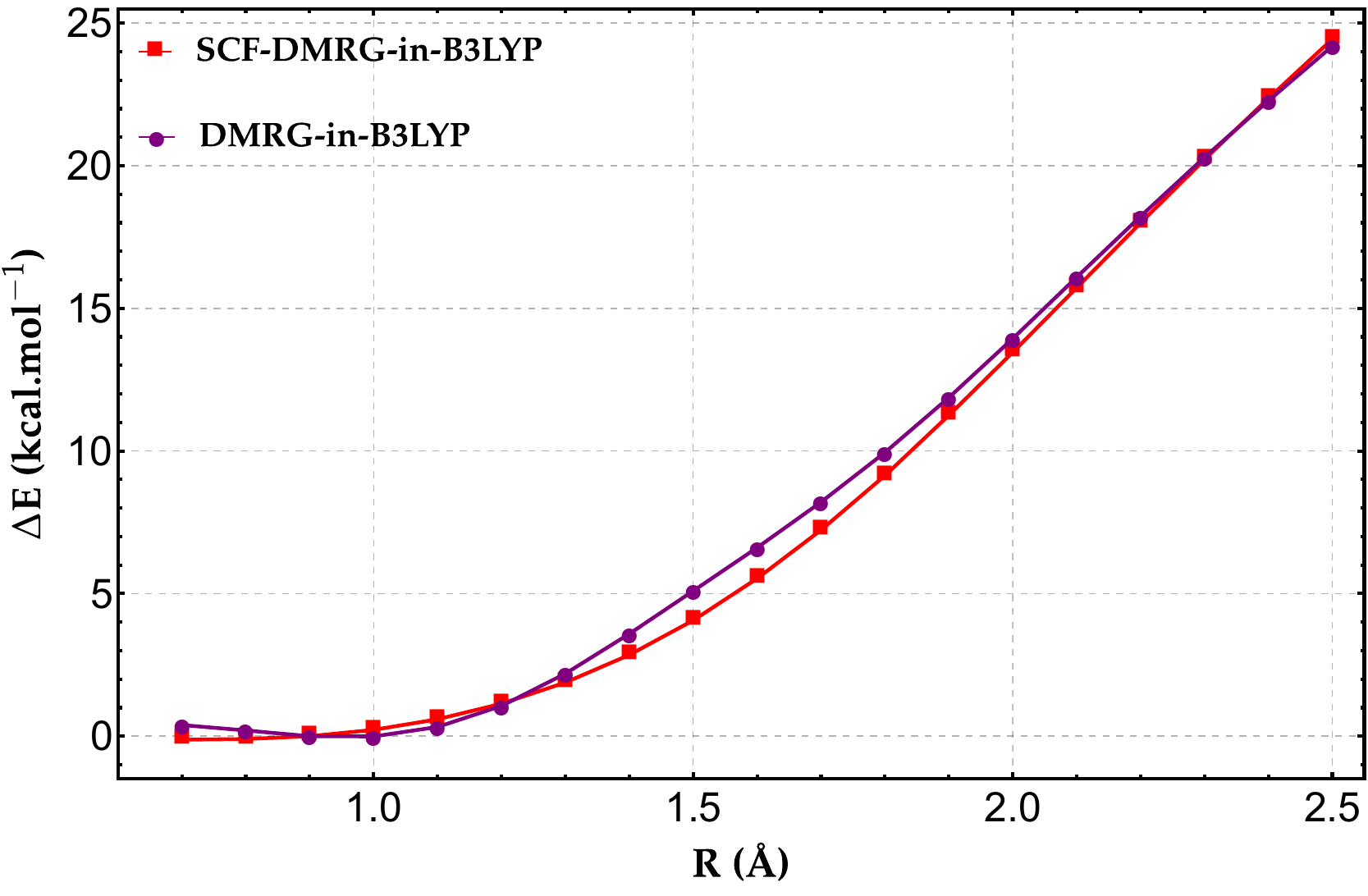}
    \caption{Relative energy error (kcal$\cdot$mol$^{-1}$) for the H$_{20}$ chain comparing self-consistent DMRG-in-B3LYP and standard DMRG-in-B3LYP, using a four-atom active fragment. Errors are referenced to full-system DMRG energies with the 6-31G basis set.}
    \label{fig:scf_vs_non_scf}
\end{figure}

We begin by comparing a standard one-shot WF-in-DFT calculation to a self-consistent (SCF) WF-in-DFT scheme. In the one-shot variant, the embedding potential is built from the initial DFT density, while in the SCF variant, the WF-in-DFT calculations are run repeatedly as macro-iterations and the embedding potential is updated at each macro-iteration using the actual WF density. The latter mitigates so-called density-driven errors \cite{Pennifold2017}, which arise when the initial DFT density does not adequately describe the active subsystem. This comparison was carried out for DMRG-in-B3LYP, without truncation of the virtual space in the DMRG calculation of the active fragment, on the H$_{20}$ chain in the 6-31G basis with a four-atom active fragment. The results are shown in Figure~\ref{fig:scf_vs_non_scf}. As expected, the SCF procedure yields slightly more accurate energies, however, the maximum deviation from the one-shot result is only about 1 kcal/mol. In practice, the perturbative “trace” term in Eq.~\ref{total_E} performs very well. 

\begin{figure}[!ht]
    \centering
    \includegraphics[width=0.6\linewidth]{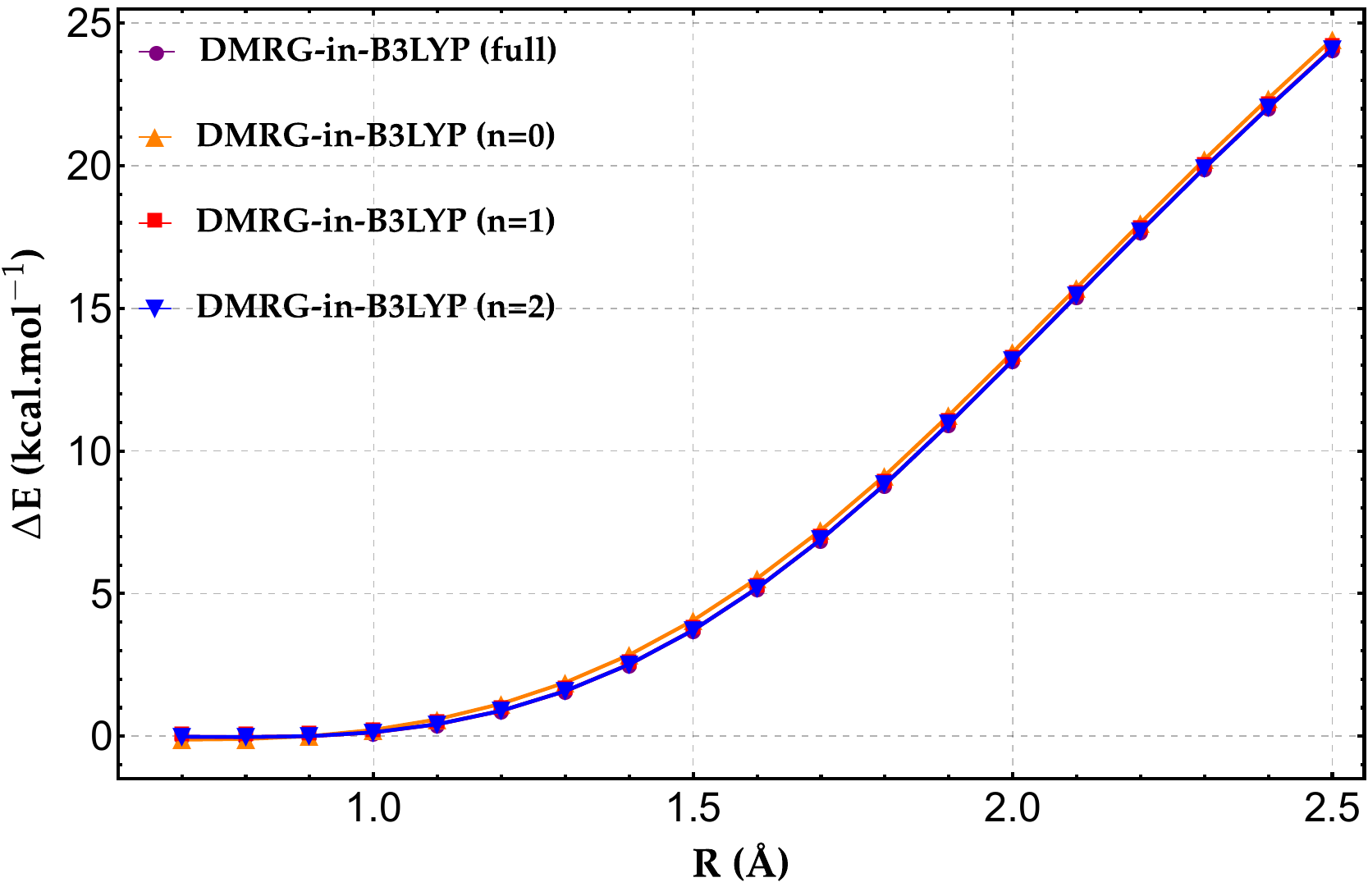}
    \caption{Relative energy error (in kcal/mol) of DMRG-in-B3LYP for the H$_{20}$ chain with 4 atoms in the active fragment, compared to full-system DMRG, using the 6-31G basis set. The labels ``full'', $n=0$, $n=1$, and $n=2$ denote the calculations without concentric localization, and with the first, second, and third CL shells, respectively.}
    \label{fig:conc_loc_compare}
\end{figure}

In Figure \ref{fig:conc_loc_compare}, we illustrate the performance of concentric localization for the H$_{20}$ chain with 4 atoms in the active fragment using the 6-31G basis.
Remarkably, sub-kcal/mol accuracy is already achieved with the first CL shell ($n = 0$), corresponding to carrying out a DMRG calculation for 4 electrons in a set of 10 orbitals, which corresponds to an active space of CAS(4,10). This is significantly smaller than the untruncated active space, CAS(4,32). Including a second CL shell ($n = 1$) fully reproduces the non-truncated results. 

Based on the observations in the 6-31G basis, we employed the one-shell CL 
approximation for subsequent calculations in the larger cc-pVDZ basis. Moreover, to eliminate any possible density-driven errors, we employ the SCF-WF-in-DFT approach (the abbreviation SCF is not explicitly specified from now on).

\begin{figure}[!ht]
    \centering
    \subfloat[\label{h4_dmrg}]{%
    \includegraphics[width=0.45\linewidth]{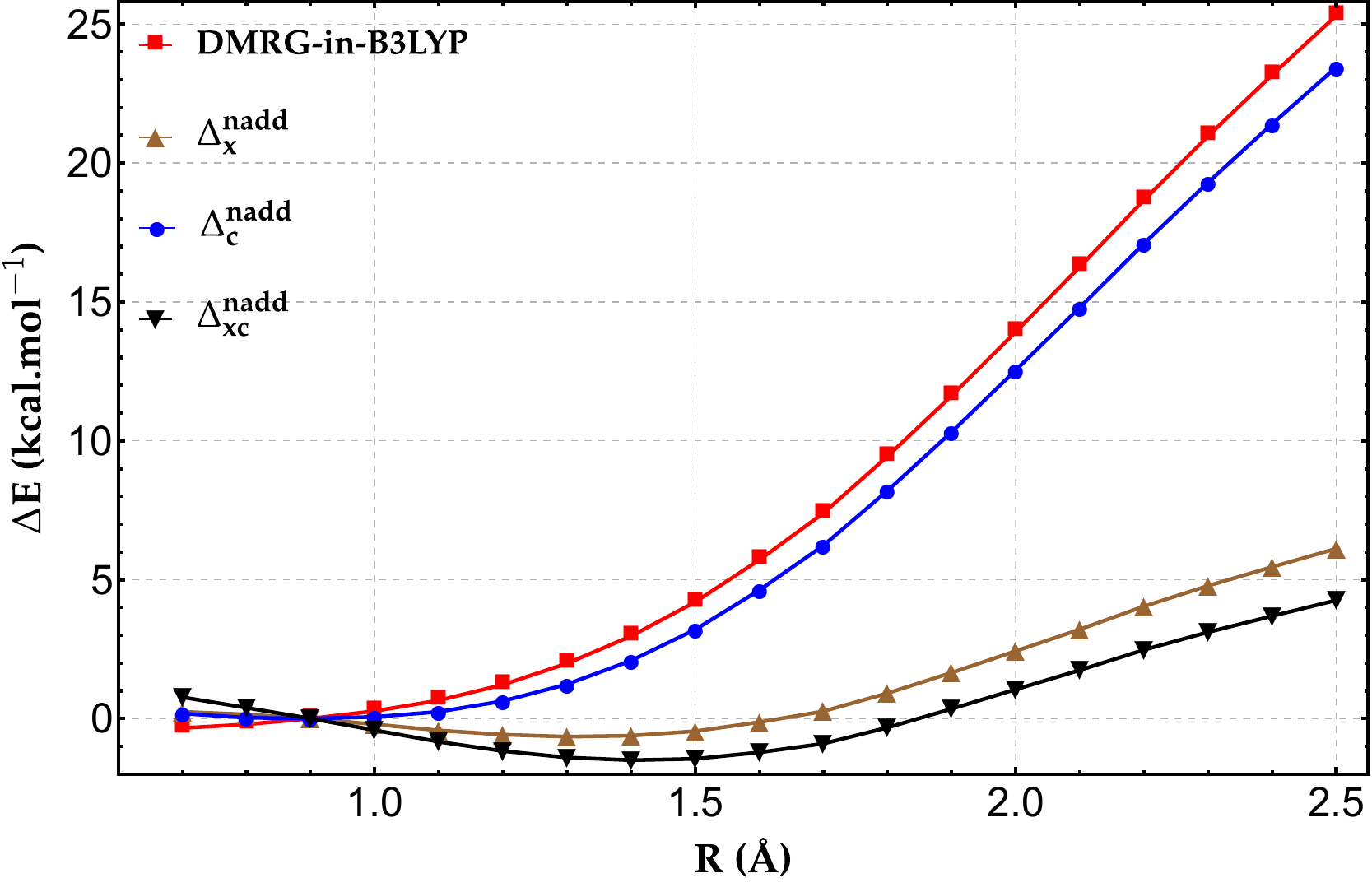}
    }
    \hskip 0.4cm
    \subfloat[\label{h8_dmrg}]{%
    \includegraphics[width=0.45\linewidth]{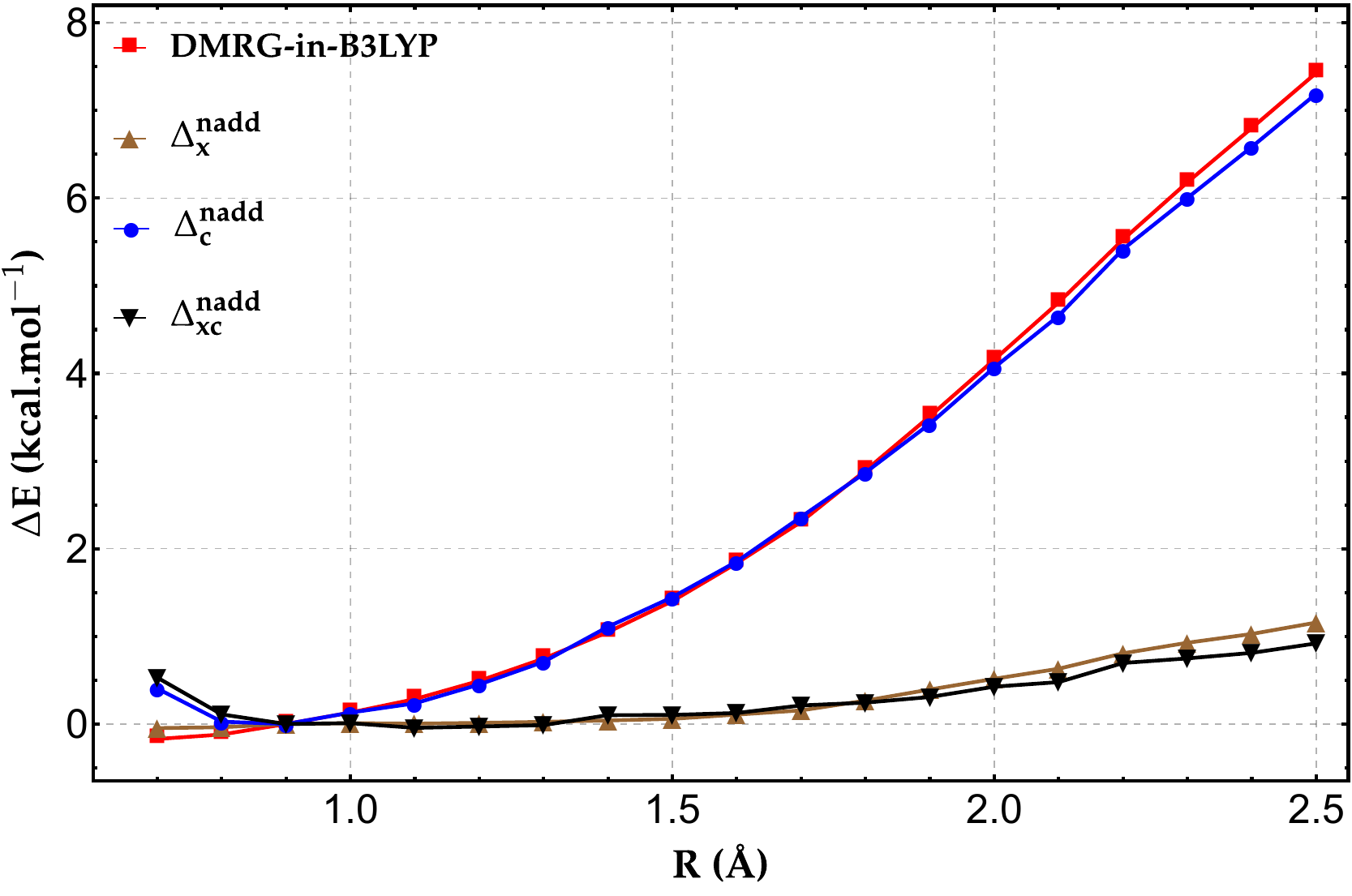}
    }
    \caption{Relative energy errors (in kcal/mol) of DMRG-in-B3LYP and DMRG-in-B3LYP with the non-additive exchange ($\Delta_x^{\text{nadd}}$), correlation ($\Delta_c^{\text{nadd}}$), and exchange–correlation ($\Delta_{xc}^{\text{nadd}}$) corrections, for H$_{20}$ chain with (a) 4-atom active fragment and (b) 8-atom active fragment, benchmarked against full-system DMRG with the cc-pVDZ basis set.}
    \label{fig:h4_h8_dmrg}
\end{figure}

We now assess the accuracy of DMRG-in-B3LYP/cc-pVDZ, both without and with non-additive exchange–correlation corrections, for the H$_{20}$ cases with 4 atoms in the active fragment (Figure \ref{h4_dmrg}) and 8 atoms in the active fragment (Figure \ref{h8_dmrg}). As shown in Figure \ref{h4_dmrg}, when the active fragment is restricted to four atoms, the energetic error of the parent DMRG-in-B3LYP method rises to nearly 25 kcal/mol for the most stretched bonds, reflecting the well-known tendency of approximate DFT functionals  to accumulate large static correlation errors in strongly correlated systems. 
As shown in Figure S1 in the Supplementary Information (SI), the errors in the total energies obtained with bare B3LYP and PBE functionals exceed 50 kcal/mol and 35 kcal/mol, respectively, in the dissociation limit of the central H$_4$ fragment. This large static-correlation error in the hydrogen chain can be primarily attributed to the fractional-spin error\cite{cohen2008fractional}, as confirmed by Figure S4. The latter shows that removing the fractional-spin error from the PBE relative energies of the entire hydrogen chain yields a dissociation curve in excellent agreement with the reference data. Most of this error arises from the exchange functional, while the correlation contribution accounts for only about 20\% of the total error.

The error extends over the central H$_4$ fragment and even beyond the first pair of hydrogen dimers adjacent to it. This indicates that correcting the static-correlation error solely within the H$_4$ or even H$_8$ fragment, by treating it with DMRG as in DMRG-in-DFT, does not suffice to recover accurate total energies for the full system. In other words, in the systems studied here, fragment A (whether H$_4$ or H$_8$) remains strongly coupled to its environment, and the DFT treatment of the nonadditive exchange–correlation energy fails to account for this coupling.

Incorporating the non-additive exact exchange together with the AC0 correlation correction significantly suppresses the error of the $E_{xc}^{\text{nadd}}$ energy, reducing it to about 4 kcal/mol, with the improvement being dominated by the exchange contribution. This is  consistent with the observation, cf.\ Figure S4 in SI, that the contribution to the total fractional spin error from the correlation functional is almost an order of magnitude smaller than that of the exchange functional (for details, see SI).
Increasing the size of the active fragment by including one adjacent hydrogen dimer on each side of the H$_4$ chain in subsystem~A  reduces the coupling between the subsystems, as most of the static-correlation effects are now captured within~A. Consequently, as shown in Figure~\ref{h8_dmrg}, the maximum error in the DMRG-in-B3LYP relative energy is much smaller than in the previous case, amounting to less than 8~kcal/mol.
With the non-additive exchange–correlation correction it drops further to less than 1~kcal/mol, here almost entirely due to the exchange term as the magnitude of the nonadditive correlation energy is much smaller. In Supporting Information we show that DMRG-in-PBE is as accurate as the B3LYP-based embedding (see Figures~S2 and S3), which indicates that hybrids may in practice be replaced with a less expensive, pure GGA model.

\begin{figure}[!ht]
    \centering
    \subfloat[\label{h4_cas}]{%
    \includegraphics[width=0.45\linewidth]{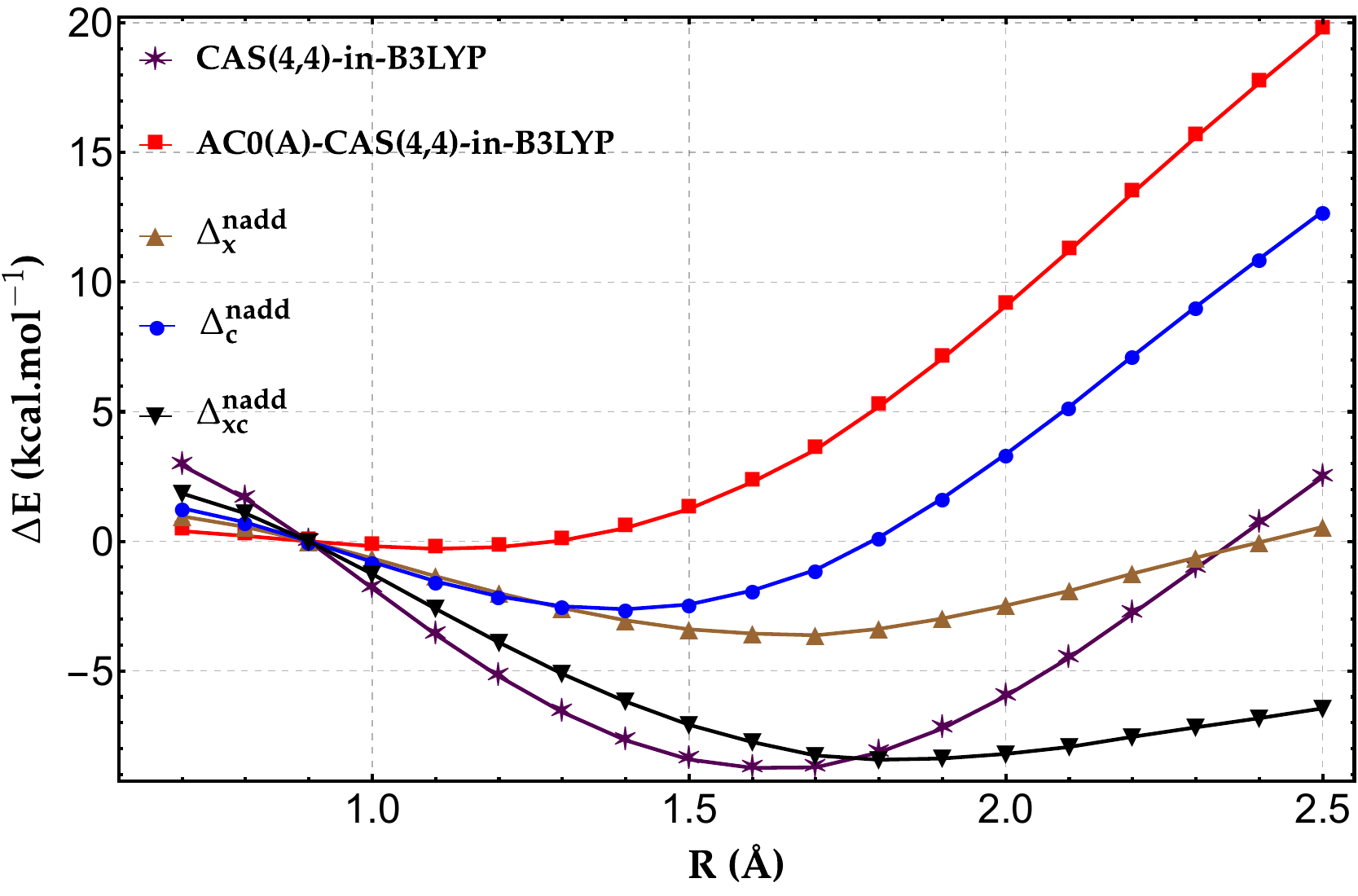}
    }
    \hskip 0.4cm
    \subfloat[\label{h8_cas}]{%
    \includegraphics[width=0.45\linewidth]{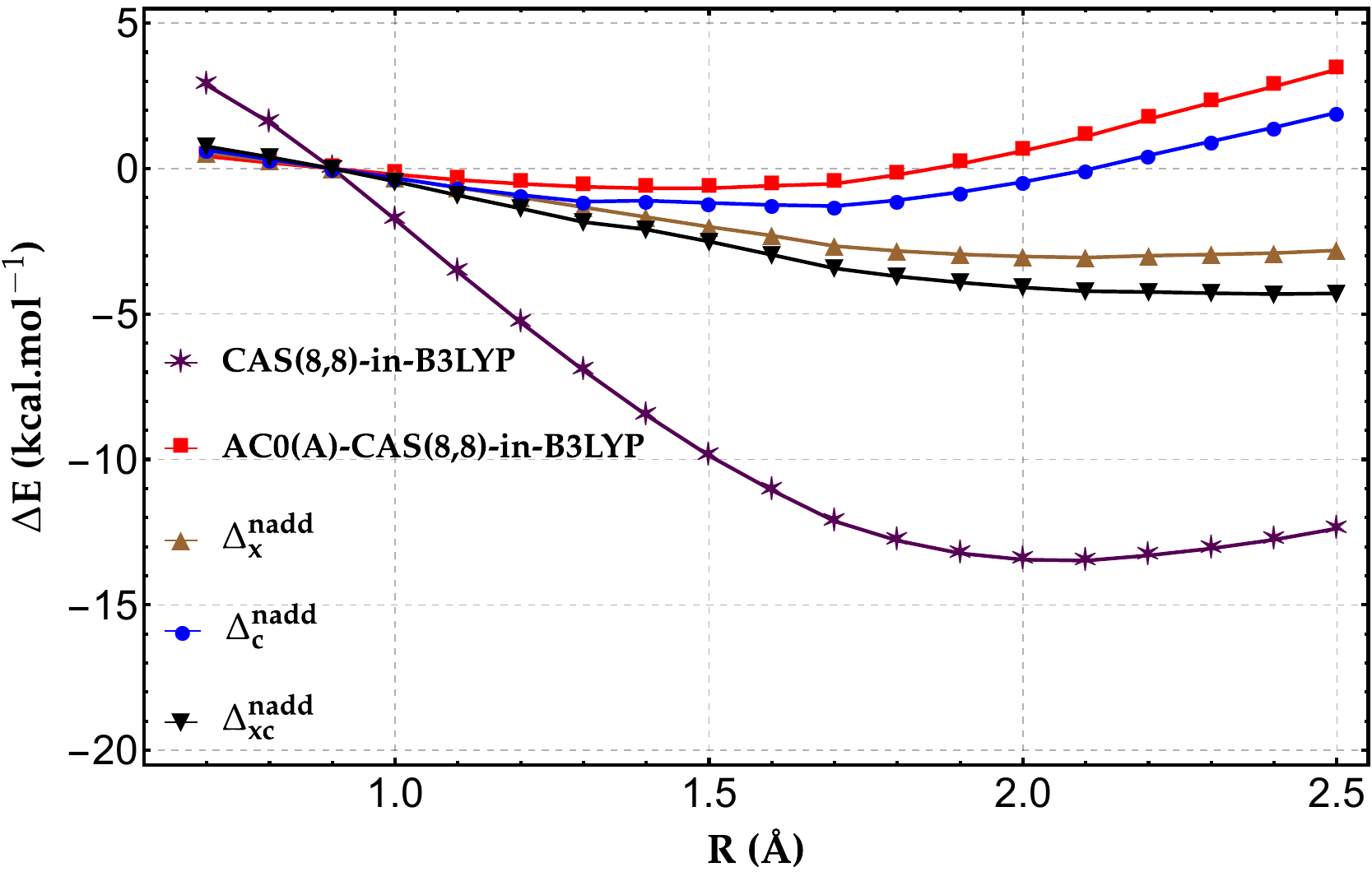}
    }
    \caption{Relative energy errors (in kcal/mol) of CAS-in-B3LYP, AC0(A)-CAS-in-B3LYP, and AC0(A)-CAS-in-B3LYP with the non-additive exchange ($\Delta_x^{\text{nadd}}$), correlation ($\Delta_c^{\text{nadd}}$), and exchange–correlation ($\Delta_{xc}^{\text{nadd}}$) corrections, for H$_{20}$ chain with (a) 4-atom active fragment and (b) 8-atom active fragment, benchmarked against full-system DMRG with the cc-pVDZ basis set.}
    \label{fig:h4_h8_cas}
\end{figure}

Since DMRG-in-DFT, even with CL, can become prohibitively expensive for larger systems due to the size of the virtual space, we also examined the performance of CAS-in-DFT. Figure \ref{fig:h4_h8_cas} shows CAS-in-B3LYP/cc-pVDZ results for the H$_{20}$ chain with (a) four atoms and (b) eight atoms in the active fragment. While CAS-in-B3LYP provides a reasonable description, it lacks dynamic electron correlation, which is often essential for achieving chemical accuracy. 
A comparison of the maximal errors obtained with CAS-in-DFT and DMRG-in-DFT for the case of a four-atom active fragment may give the false impression that CAS-in-DFT is more accurate. In fact, this apparent improvement arises from a fortuitous cancellation between the negative error in the CASSCF relative energy, caused by the missing dynamic correlation, and the positive error in the nonadditive exchange–correlation contribution. When the active fragment is enlarged to eight atoms, this favorable cancellation disappears, and the CAS-in-DFT results become less accurate than those of DMRG-in-DFT. Hence, achieving consistent accuracy within CAS-in-DFT requires simultaneous correction of both the CAS and nonadditive exchange–correlation energies.
To address this, we applied the AC0 correction to first recover the missing dynamical correlation within the active subsystem. We denote this method as AC0(A)-CAS-in-B3LYP. 
The largest errors for CAS-in-B3LYP in the four-atom case occur at intermediate bond stretches. Once the AC0 correction is included, the resulting error profile changes and closely resembles that of DMRG-in-B3LYP (see Figure \ref{h4_dmrg}) with the largest errors for the most stretched bonds. This picture is consistent with the fact that AC0 captures the dominant part of dynamic electron correlation, thereby shifting the CAS-in-B3LYP results toward DMRG-in-B3LYP. 
A similar effect is observed for the eight-atom active fragment: describing it solely at the CASSCF level leads to a substantially negative error in the relative energy, whereas inclusion of dynamical correlation for the active fragment brings the CAS-in-B3LYP results into close agreement with DMRG-in-B3LYP.

Next, we focus on assessing the impact of the non-additive exchange–correlation corrections in the case of CAS-in-B3LYP. Adding the nonadditive exchange energy correction to AC0(A)-CAS-in-B3LYP results in the same effect as in DMRG-in-B3LYP, reducing the maximal relative energy error by 20 kcal/mol and 7 kcal/mol for the four- and eight-atom active fragment partitions, respectively.

As presented in the Theory section, the nonadditive correlation energy correction for CAS-in-DFT is derived under the same assumption as for DMRG-in-DFT—namely, that only the AC0 amplitudes which vanish in the limit of infinite separation between the active fragment and the environment are retained in the correlation energy expression. Nevertheless, the numerical values of the resulting corrections are not expected to coincide for the two methods.
The first important observation is that the chosen active space in CAS-in-DFT consists of orbitals exclusively localized on the active fragment in the separation limit. This has important implications for the contributions to the non-additive correlation energy, as compared to the DMRG-in-DFT case. Since no active orbitals will be localized on the environment fragment when the fragments are infinitely separated, we cannot apply the same exclusions as in the latter method. In particular, the terms containing AC0 amplitudes $T^{\text{AC0}}_{\text{(ao)(ao)}}$ and $T^{\text{AC0}}_{\text{(vo)(ao)}}$, satisfy Eq.~(\ref{cond1}) and must now be retained in the nonadditive correlation energy formula. In fact, all terms that mix indices from the occupied (o), active (a), and virtual (v) spaces contribute to the latter.
Another difference arises from the treatment of the virtual space. Unlike in DMRG-in-DFT, in CAS-in-DFT we do not employ the CL and truncation of the virtual space. As a consequence, the virtual orbitals do correlate with the active orbitals and  $T^{\text{AC0}}_{\text{(va)(..)}}$ AC0 amplitudes remain finite, which further enhances the overall contribution of the non-additive correlation energy. This trend is clearly reflected in the results presented in Figures \ref{h4_cas} and \ref{h8_cas}. The inclusion of the nonadditive correlation correction yields a noticeable improvement in AC0(A)-CAS-in-B3LYP. However, adding the nonadditive exchange contribution on top leads to an underestimation of the relative energy. Comparable accuracy is obtained for four- and eight-atom active fragments, with errors at the largest bond elongations near $-5$ kcal/mol. 

\begin{figure}[!ht]
    \centering
    \includegraphics[width=0.6\linewidth]{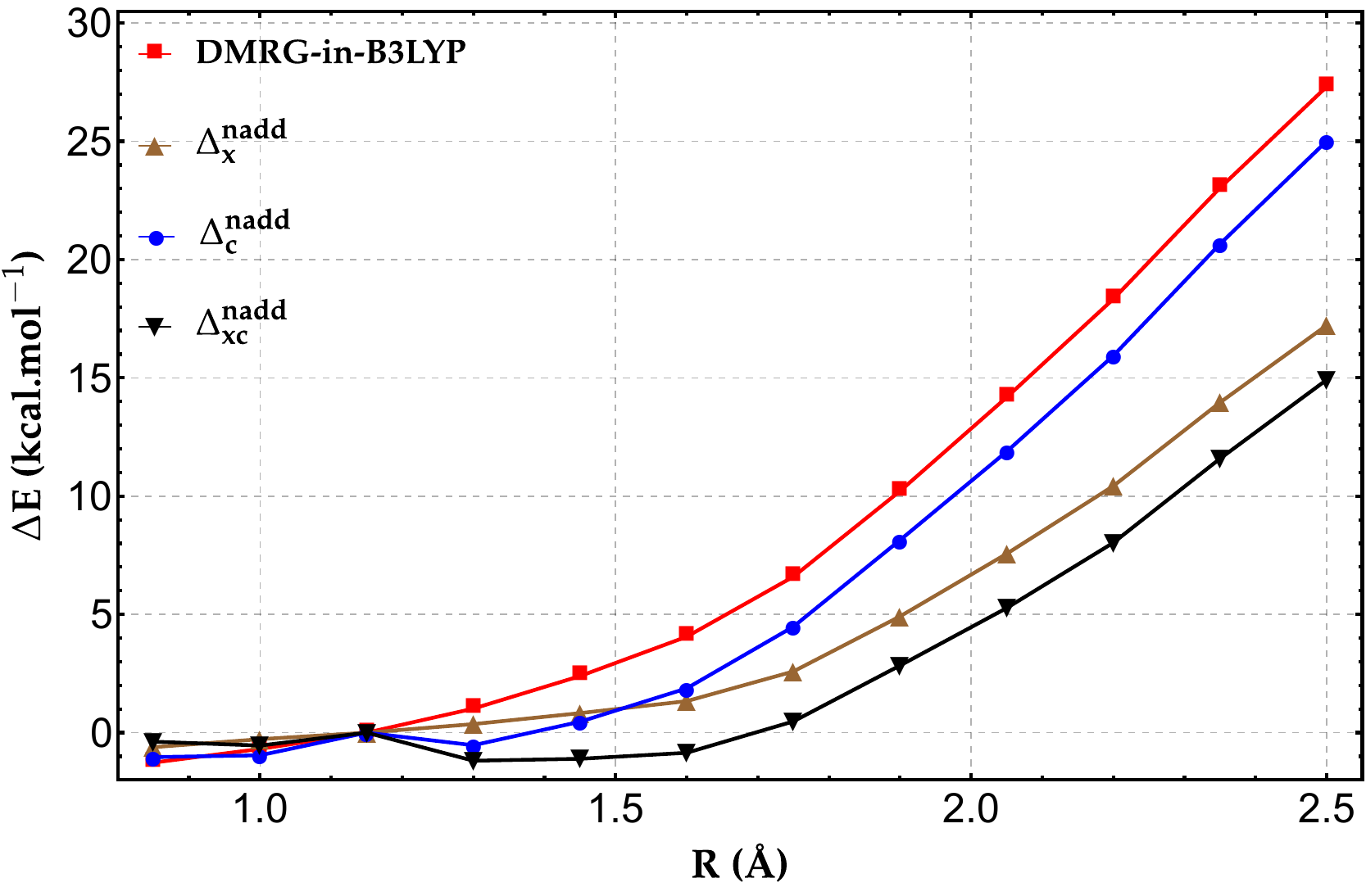}
    \caption{Relative energy errors (in kcal/mol) of DMRG-in-B3LYP and DMRG-in-B3LYP with the non-additive exchange ($\Delta_x^{\text{nadd}}$), correlation ($\Delta_c^{\text{nadd}}$), and exchange–correlation ($\Delta_{xc}^{\text{nadd}}$) corrections, for the triple bond stretching in propionitrile (CH$_3$CH$_2$CN) benchmarked against full-system frozen-core DMRG with the cc-pVDZ basis set.}
    \label{fig:propio_dmrg}
\end{figure}

\begin{figure}[!ht]
    \centering
    \includegraphics[width=0.6\linewidth]{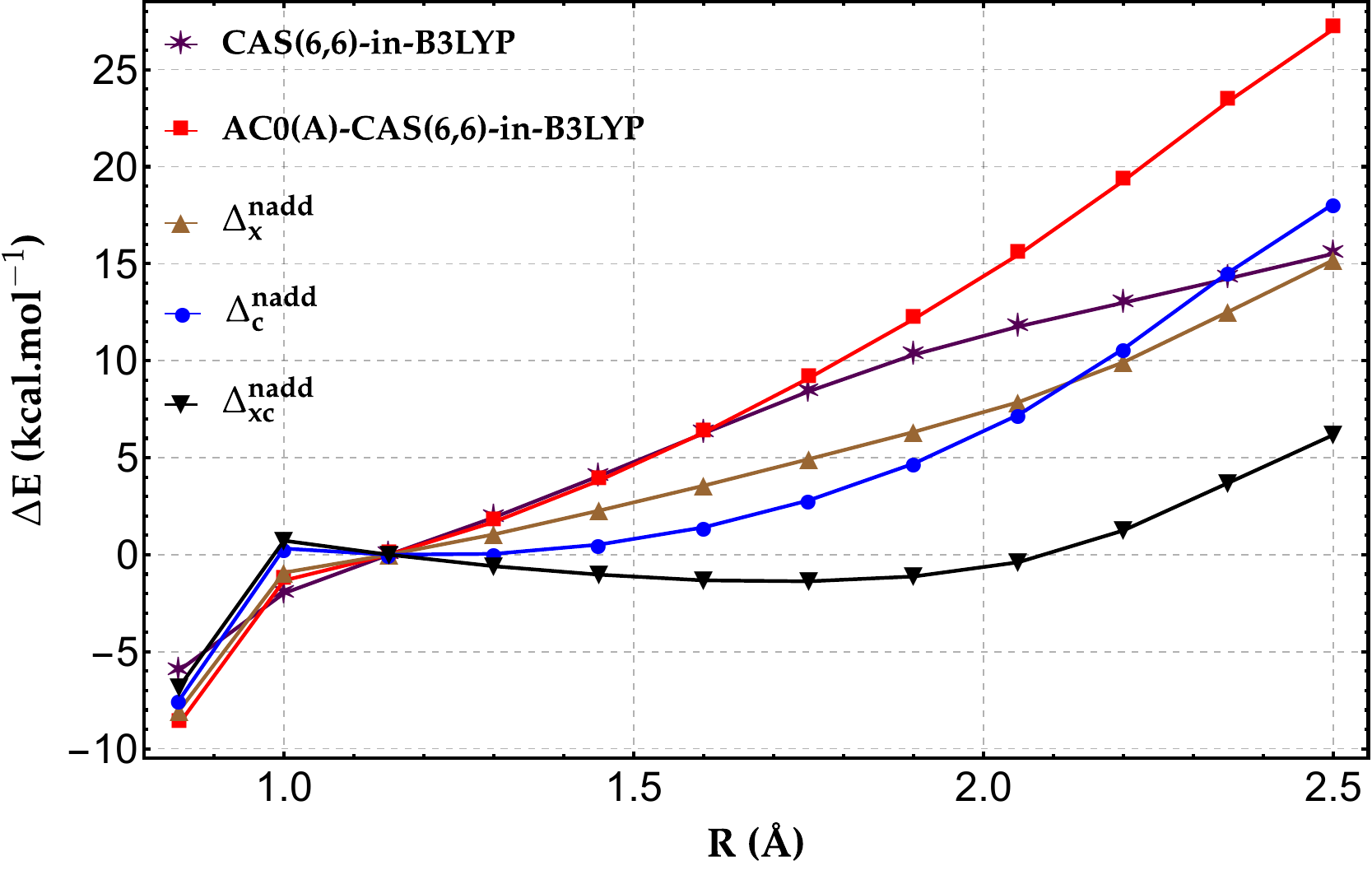}
    \caption{Relative energy errors (in kcal/mol) of CAS(6,6)-in-B3LYP, AC0(A)-CAS(6,6)-in-B3LYP, and AC0(A)-CAS(6,6)-in-B3LYP with the non-additive exchange ($\Delta_x^{\text{nadd}}$), correlation ($\Delta_c^{\text{nadd}}$), and exchange–correlation ($\Delta_{xc}^{\text{nadd}}$) corrections, for the triple bond stretching in propionitrile (CH$_3$CH$_2$CN) benchmarked against full-system frozen-core DMRG with the cc-pVDZ basis set.}
    \label{fig:propio_cas}
\end{figure}

\subsection{Propionitrile}
The second system we examined is the triple-bond stretching in propionitrile. The corresponding results are presented in Figures \ref{fig:propio_dmrg} and \ref{fig:propio_cas} for DMRG-in-B3LYP/cc-pVDZ and CAS-in-B3LYP/cc-pVDZ, respectively. In the case of DMRG-in-B3LYP, we observe a similar trend to that found for the H$_{20}$ chain: the non-additive exchange correction has a stronger influence than the non-additive correlation correction. 
However, its overall effect is smaller than in the hydrogen chain, resulting in a larger residual error in the corrected DMRG-in-B3LYP energies. For the most stretched bond, the remaining error exceeds 10 kcal/mol, which nevertheless represents a substantial improvement, by more than 10 kcal/mol, relative to the uncorrected DMRG-in-DFT results.

By contrast, the corrected AC0(A)-CAS(6,6)-in-B3LYP results remain within $\pm 5$ kcal/mol for most bond lengths, except for the first and last points. As in the H$_{20}$ chain, the uncorrected CAS(6,6)-in-B3LYP performs slightly better than AC0(A)-CAS(6,6)-in-B3LYP due to error cancellation. We also find that the nonadditive correlation correction plays a more prominent role in AC0(A)-CAS(6,6)-in-B3LYP than in DMRG-in-B3LYP, accounting for the overall lower errors of the former method compared to the latter. This behavior can be attributed to the fact that the nonadditive correlation term in CAS-in-DFT contains a larger number of contributing amplitudes (cf.\ Table~\ref{tab:corr}) than its counterpart in DMRG-in-DFT.

\section{Conclusions}
\label{section_conclusions}
The projection-based DMRG-in-DFT method offers a promising framework for systems in which strong correlation effects are present and localized within a relatively small region of the total system, allowing the DFT description to be replaced by that of a multireference wavefunction method. 
However, including all atoms that contribute significantly to static correlation would require a prohibitively large active space, so in practice the wavefunction fragment remains strongly coupled to its environment. As demonstrated in Ref.~\citenum{Beran2023}, the coupling gives rise to substantial errors in DMRG-in-DFT embedding. In this work, we corroborated this finding by reformulating the total energy expression of the composite system to isolate the nonadditive exchange–correlation energy term, which allowed us to identify it as the principal source of error. To overcome the inability of standard DFT functionals to capture the strong coupling between the active fragment and its environment, we introduced physically motivated corrections for the nonadditive exchange and correlation energies.
Specifically, the nonadditive exchange energy is evaluated using the exact exchange functional, while the nonadditive correlation energy is derived from the AC0 correlation energy, retaining only those contributions that vanish in the limit of infinite separation between the fragment and its environment. The new approach enables accurate predictions even with relatively small active fragments.


To demonstrate the reliability of the proposed corrections, we chose two model systems for which DFT fails due to the pronounced static correlation error: a linear H$_{20}$ chain with three central covalent bonds simultaneously stretched, and the propionitrile molecule undergoing dissociation of the \ce{C#N} triple bond.
While the uncorrected DMRG-in-DFT approach already offers a substantial improvement over pure DFT, the relative energy errors remain considerable—exceeding 25 kcal/mol for the hydrogen chain with a four-atom active fragment and for propionitrile. Upon applying the exchange–correlation energy corrections, the error is significantly reduced in all studied systems.


In practical applications of the projection based DMRG-in-DFT embedding to systems with large environments, evaluating the nonadditive AC0 correlation correction may become prohibitively expensive because of the large number of inactive orbitals assigned to the environment and the extensive virtual space involved (the computational cost of AC0 scales with the fifth power of the number of orbitals). In such cases, one can take advantage of the locality of the environmental orbitals and truncate the ERPA matrices used to compute the AC0 correlation amplitudes, retaining only those orbitals that lie in close proximity to the active fragment. Similar truncation strategy could be adopted for the virtual orbitals upon their localisation. 
An even more efficient alternative is to employ fractional-spin–error-free functionals for evaluating the nonadditive exchange–correlation energy. Work in these directions is currently in progress.

\section*{Supplementary Information}
Full KS-DFT (with B3LYP and PBE functionals) curves for the H$_{20}$ hydrogen chain and propionitrile (CH$_3$CH$_2$CN) molecule. DMRG-in-PBE for 4-atom and 8-atom active fragment for the H$_{20}$ hydrogen chain. Absolute energies of the H$_{20}$ hydrogen chain for full KS-PBE, DMRG-in-B3LYP, CAS-in-B3LYP and full DMRG. Full KS-PBE curve and dissociation energy corrected for the fractional spin error for the H$_{20}$ hydrogen chain. Equilibrium geometry and absolute energies for KS-B3LYP, DMRG-in-B3LYP, CAS-in-B3LYP and full DMRG for the propionitrile molecule.

\section*{Acknowledgments}

This work was supported by the Czech Science Foundation (Grant No. 23-04302L); the National Science Center of Poland (Grant No. 2021/43/I/ST4/02250); the Czech Ministry of Education, Youth and Sports through the e-INFRA CZ (ID:90254) and the Advanced Multiscale Materials for Key Enabling Technologies project, Project No. CZ.02.01.01/00/22\_008/ 0004558, Co-funded by the European Union; and the Center for Scalable and Predictive methods for Excitation and Correlated phenomena (SPEC), which is funded by the U.S. Department of Energy (DOE), Office of Science, Office of Basic Energy Sciences, the Division of Chemical Sciences, Geosciences, and Biosciences.

\section{Appendix}
The computation of AC0 amplitudes for the nonadditive correlation energy
follows the procedure outlined in Refs.~\citenum{pastorczak2018correlation,guo2024spinless},
with one essential modification: the effective one-electron Hamiltonian,
$\hat{h}^{(0)}$, employed in the zeroth-order ERPA equations must include the
embedding potential within the active orbital subspace.

Specifically, if the embedded  $\Psi^{A}$ wavefunction follows from DMRG\ calculation, i.e.\ all orbitals assigned to A are active then the effective Hamiltonian is constructed as follows
\begin{equation}
h_{pq}^{(0)}=\left\{
\begin{array}
[c]{cc}%
h_{pq}+\upsilon_{pq}^{\text{emb}} & p,q\in\text{a}\\
h_{pq}+\left[  \text{JK}\right]  _{pq} & \ \ \ \ \ \ \ p,q\in\text{o}%
_{\text{B}}\ \ \text{or\ \ }p,q\in\text{v}\\
0 & \text{otherwise}%
\end{array}
\right.
\end{equation}
where o$_{\text{B}}$ denotes a set of the inactive orbitals assigned to B, v
stands to a set of virtual orbitals, which is not empty if DMRG\ employs
truncated space of the HF\ virtual orbitals. The Coulomb-exchange interaction
energy matrix [JK] is defined as follows%
\begin{align}
\forall_{pq\in\text{o}_{\text{B}}}\ \ \ \left[  \text{JK}\right]  _{pq} &
=\sum_{rs\in\text{a}}\left[  \gamma_{\text{emb}}^{A}\right]  _{rs}\left(
 \left\langle pr|qs\right\rangle -\left\langle pr|sq\right\rangle \right)  \\
\forall_{pq\in\text{v}}\ \ \ \left[  \text{JK}\right]  _{pq} &  =\sum
_{rs\in\text{a}}\left[  \gamma_{\text{emb}}^{A}\right]  _{rs}\left(
 \left\langle pr|qs\right\rangle -\left\langle pr|sq\right\rangle \right)
\nonumber\\
&  +\sum_{rs\in\text{o}_{\text{B}}}\left[  \gamma^{B}\right]  _{rs}\left(
 \left\langle pr|qs\right\rangle -\left\langle pr|sq\right\rangle \right)
\label{JKv}%
\end{align}

In the CAS-in-DFT embedding method, the nonadditive correlation energy
correction is derived from the AC0 amplitudes obtained using an effective
Hamiltonian that accounts for the presence of the inactive orbital set
$o_{\text{B}}$ localized on subsystem~A and reads
\begin{equation}
h_{pq}^{(0)}=\left\{
\begin{array}
[c]{cc}%
h_{pq}+\left[  \text{JK}\right]  _{pq}+\upsilon_{pq}^{\text{emb}} &
p,q\in\text{a}\\
h_{pq}+\left[  \text{JK}\right]  _{pq} & p,q\in\text{o}_{\text{A}%
}\ \ \text{or}\ \ p,q\in\text{o}_{\text{B}}\ \ \text{or\ \ }p,q\in\text{v}\\
0 & \text{otherwise}%
\end{array}
\right.
\end{equation}
where%
\begin{align}
\forall_{p,q\in\text{a}}\ \ \ \left[  \text{JK}\right]  _{pq}  & =\sum
_{rs\in\text{o}_{\text{A}}}\left[  \gamma_{\text{emb}}^{A}\right]
_{rs}\left(   \left\langle pr|qs\right\rangle -\left\langle pr|sq\right\rangle
\right)  \\
\forall_{p,q\in\text{o}_{\text{A}}}\ \ \ \left[  \text{JK}\right]  _{pq}  &
=\sum_{rs\in\text{a}}\left[  \gamma_{\text{emb}}^{A}\right]  _{rs}\left(
 \left\langle pr|qs\right\rangle -\left\langle pr|sq\right\rangle \right)  \\
\forall_{pq\in\text{o}_{\text{B}}}\ \ \ \left[  \text{JK}\right]  _{pq}  &
=\sum_{rs\in\text{a}}\left[  \gamma_{\text{emb}}^{A}\right]  _{rs}\left(
 \left\langle pr|qs\right\rangle -\left\langle pr|sq\right\rangle \right)  \\
& +\sum_{rs\in\text{o}_{\text{A}}}\left[  \gamma_{\text{emb}}^{A}\right]
_{rs}\left(   \left\langle pr|qs\right\rangle -\left\langle pr|sq\right\rangle
\right)
\end{align}
and the matrix $\left[  \text{JK}\right]  _{pq\in\text{v}}$ is defined in
Eq.(\ref{JKv}).

\bibliography{references}

\end{document}